\documentstyle[mnras_cite,psfig]{mn}
\def\propsim{~\rlap{\raise 0.25ex\hbox{$\propto$}}{\lower 0.75ex\hbox{$\sim$}}}
\def\gsim{~\rlap{$>$}{\lower 1.0ex\hbox{$\sim$}}}
\def\lsim{~\rlap{$<$}{\lower 1.0ex\hbox{$\sim$}}}
\def\G{{\rm G}}
\def\HI{{\rm H}{\sc i}}
\def\HeII{{\rm He}{\sc ii}}
\def\HeIII{{\rm He}{\sc iii}}
\def\d{{\rm d}}
\def\Lya{Ly$\alpha\ $}
\def\etal{{et al.~}}

\begin{document}

\title[Early Preheating and Galaxy Formation]{Early Preheating and Galaxy Formation}
\author[A.~J.~Benson \& Piero Madau]{A.~J.~Benson$^1$ \& Piero Madau$^2$\\
1. California Institute of Technology, MC 105-24, Pasadena, CA 91125,
U.S.A. (e-mail: abenson@astro.caltech.edu) \\
2. Department of Astronomy and Astrophysics, University of 
California, Santa Cruz, CA 95064, U.S.A. (e-mail: pmadau@ucolick.org)}

\maketitle

\begin{abstract}
Winds from pregalactic starbursts and `miniquasars' may pollute the
intergalactic medium (IGM) with metals and raise its temperature to a
much higher adiabat than expected from photoionization, and so inhibit
the formation of early galaxies by increasing the cosmological Jeans
mass. We compute the thermal history of the IGM when it experiences a
period of rapid, homogeneous ``preheating'' at high redshifts, and the
impact of such a global feedback mechanism on the IGM ionization state
and the subsequent galaxy formation and evolution. Measurements of the
temperature of the \Lya forest at redshift $z\sim 3$ constrain the
redshift and energy of preheating, and rule out models that preheat
too late or to too high a temperature, i.e. to $T_{\rm IGM}\gsim
10^6\,$K at $z\lsim 10$. The IGM thermal history is used to estimate
the effects of preheating on the formation of galaxies at later
epochs, allowing us to predict galaxy luminosity functions in
preheated universes. The results depend crucially on whether the
baryonic smoothing scale in the IGM is computed globally, or in a
local, density-dependent fashion (since the IGM temperature can become
highly inhomogeneous in the post-preheating epoch). Using a globally
averaged smoothing scale, we find that models with excessive
preheating produce too few $L_*$ and fainter galaxies, and are
therefore inconsistent with observational data. More moderate
preheating scenarios, with $T_{\rm IGM}\gsim 10^5\,$K at $z\sim 10$,
are able to flatten the faint-end slope of the luminosity function,
producing excellent agreement with observations, without the need for
any local feedback mechanism within galaxies. A density-dependent
smoothing scale requires more energetic preheating to achieve the same
degree of suppression in the faint-end slope. All models, however,
appear unable to explain the sharp cut-off in the luminosity function
at bright magnitudes---a problem that is also common to more
conventional local feedback prescriptions.  Supernova-driven
preheating scenarios tend to raise the mean metallicity of the
universe well above the minimum levels observed in the \Lya
clouds. The high energies associated with preheating cause a sharp
drop in the abundance of neutral hydrogen in the IGM and are often
sufficient to double ionize helium at high redshift, well before the
`quasar epoch'. We find that ionizing photon escape fractions must be
significantly higher than 10\% in order to explain the low inferred
\HI\ fraction at $z\approx 6$, particularly when using a globally
averaged smoothing scale. While early preheating causes strong
suppression of dwarf galaxy formation we show that it is not able to
reproduce the observed abundance of satellite galaxies in the Local
Group in detail. The detailed thermal history of the universe during
the formative early stages around $z=10-15$ remains one of the crucial
missing links in galaxy formation and evolution studies.\\

\end{abstract}

\section{Introduction}

In cold dark matter (CDM) cosmological scenarios, structure formation
is a hierarchical process in which non-linear, massive structures grow
through merging of smaller initial units. Large numbers of low-mass
dark halos are predicted to be present at early times in these popular
theories, and galaxies are thought to form by a two-stage collapse
process: the gas first infalls along with the dark matter
perturbation, gets shock-heated to the virial temperature, condenses
rapidly due to atomic or molecular line cooling, and then becomes
self-gravitating (but see \pcite{katz02} and \pcite{bd03} for
alternative views of how gas reaches the galaxy phase). Massive stars
subsequently form with some initial mass function (IMF), synthesize
heavy-elements, and explode as supernovae (SNe) after $\sim 10^7\,$yr,
enriching the surrounding medium. The very first zero-metallicity
stars (`Population III') may in fact have been so massive to give
origin to a numerous population of massive `seed' black holes (Madau
\& Rees 2001).

It is an early generation of subgalactic stellar systems around a
redshift of 10--15, aided by a population of accreting black holes in
their nuclei, which likely generated the ultraviolet radiation and
mechanical energy that ended the cosmic ``dark ages'' and reheated and
reionized most of the hydrogen in the intergalactic medium (IGM). The
recent analysis of the first year data from the {\it Wilkinson
Microwave Anisotropy Probe} ({\it WMAP}) satellite suggests the
universe was reionized at redshift $z_{\rm ion}=20^{+10}_{-9}$ (Kogut
\etal 2003).

The detailed history of the universe during and soon after these
crucial formative stages depends on the power-spectrum of density
fluctuations on small scales and on a complex network of poorly
understood `feedback' mechanisms. Yet, it is a simple expectation of
the above scenario that the energy deposition by SN explosions and
winds from accreting black holes (termed `miniquasars' in Haiman,
Madau \& Loeb 1999) in the shallow potential wells of subgalactic
systems may, depending on the efficiency with which halo gas can cool
and fragment into clouds and then into massive stars and black holes,
cause the blow-away of metal-enriched baryons from the host galaxy and
the pollution of the IGM at early times (e.g. Dekel \& Silk 1986;
Tegmark, Silk \& Evrard 1993; Cen \& Ostriker 1999; Madau, Ferrara \&
Rees 2001; Aguirre \etal 2001).

It has long been argued that, besides being a mechanism for spreading
metals around, pregalactic outflows must also efficiently quench high
redshift star formation. This is because the cooling time of
collisionally ionized high density gas in subgalactic systems is much
shorter than the then Hubble time, virtually all baryons are predicted
to sink to the centres of these small halos in the absence of any
countervailing effect (White \& Rees 1978). Strong feedback is then
necessary in hierarchical clustering scenarios to avoid this `cooling
catastrophe', i.e. to prevent too many baryons from turning into stars
as soon as the first levels of the hierarchy collapse.  The required
reduction of the stellar birthrate in halos with low circular
velocities may naturally result from the heating and expulsion of
material due to quasar winds and repeated SN explosions from an early
burst of star formation.

It has also been recognized that the radiative and mechanical energy
deposited by massive stars and accreting black holes into the
interstellar medium of protogalaxies may have a more global negative
feedback on galaxy formation. The photoionizing background responsible
for reionizing the IGM will both increase gas pressure support
preventing it from collapsing into low-mass halos along with the dark
matter, and reduce the rate of radiative cooling of gas inside halos
(Efstathiou 1992; Thoul \& Weinberg 1996; Navarro \& Steinmetz 1997).
Furthermore, as the blast-waves produced by miniquasars and
protogalaxies propagate into intergalactic space, they may drive vast
portions of the IGM to a much higher adiabat than expected from
photoionization (e.g. Voit 1996; Madau 2000; Madau \etal 2001; Theuns,
Mo \& Schaye 2001; Cen \& Bryan 2001), so as to `choke off' the
collapse of further galaxy-scale systems by raising the cosmological
Jeans mass. The Press-Schechter theory for the evolving mass function
of dark matter halos predicts a power--law dependence, $\d N/\d\ln
m\propto m^{(n_{\rm eff}-3)/6}$, where $n_{\rm eff}$ is the effective
slope of the CDM power spectrum, $n_{\rm eff}\approx -2.5$ on
subgalactic scales. As hot outflowing gas escapes its host halo,
shocks the IGM, and eventually forms a blast wave, it sweeps a region
of intergalactic space the volume of which increases with the $3/5$
power of the injected energy $E_0$ (in the adiabatic Sedov-Taylor
phase). The total fractional volume or porosity, $Q$, filled by these
hot bubbles is then $ Q\propto E_0^{3/5}\,\d N/\d\ln m$. The
dependence of $E_0$ on halo mass is unknown and depends upon the
complex physics of star formation occurring with each halo. For
illustrative purposes we will assume that the energy per logarithmic
mass interval is constant, $E_0 \d N/\d\ln m=$constant (which, for the
scales of interest, results in $E_0 \propsim m$). In this case we
find, $Q \propto (\d N/\d\ln m)^{2/5}\propto m^{-11/30}$.  Within this
simple picture it is the halos with the smallest masses which will
arguably be the most efficient at heating the IGM on large scales (to
avoid this would require $E_0\propto m^\alpha$ with $\alpha \gsim
1.5$). Note that this type of global feedback is fundamentally
different from the `in situ' heat deposition commonly adopted in
galaxy formation models, in which hot gas is produced by supernovae
within the parent galaxy.  In the following we will refer to this
global early energy input as ``preheating''. A large scale feedback
mechanism may also be operating in the intracluster medium: studies of
X-ray emitting gas in clusters show evidence for some form of
non-gravitational entropy input \cite{ponmanetal}. The energy required
is at a level $\sim 1\,$keV per particle, and must be injected either
in a more localized fashion or at late epochs in order not to violate
observational constraints on the temperature of the \Lya forest at
$z\sim 3$ (see below). Of course, since this is sufficient to
substantially alter the distribution of gas in cluster-sized
potentials, it will have a much larger effect on gas in galaxy-sized
potentials.

Preheating by definition causes a large increase in the temperature of
the IGM at high redshift. This consequently increases the Jeans mass,
thereby preventing gas accreting efficiently into small dark matter
halos. If the Jeans mass is sufficiently high bright galaxies will not
be able to form, resulting in an inconsistency with the observed
galaxy luminosity function. For typical preheating energies (see
\S\ref{sec:results}) the IGM is expected to be driven to temperatures
just below the virial temperatures of halos hosting $L_*$
galaxies. Thus we may expect preheating to have a strong effect on the
galaxy luminosity function at $z=0$.  Recently \scite{momao} and
\scite{ohajb} have studied the effects of `late preheating' on the
formation of galaxies at lower redshifts, finding that this may have a
strong impact on both the abundances and morphologies of
galaxies. Here, we perform a detailed calculation of the effect of a
global energy input in the IGM at the end of the cosmic dark ages.  By
computing the thermal history of a preheated universe we are able to
constrain both the amount and epoch of energy deposition.  While
explosion-driven winds may also inhibit the formation of nearby
low-mass galaxies through other processes, such as `baryonic
stripping' (e.g. Scannapieco, Ferrara \& Madau 2002), in this work we
assess the effect of the increased gas pressure after preheating on
subsequent galaxy formation, and use the techniques of
\scite{benson02a} to compute the resulting luminosity functions of
galaxies.\footnote{Note that similar techniques were developed by
\protect\scite{shapiro94}, although they were not employed to compute
the galaxy luminosity function.} Specifically, we investigate what
constraints the observed galaxy luminosity function (LF) and inferred
\HI\ fractions at $z\approx 6$ place on preheating scenarios and ask
whether an early homogeneous heat deposition in the IGM may provide
sufficient suppression of galaxy formation to explain the very flat
faint end slope of the LF.

The remainder of this paper is arranged as follows. In
\S\ref{sec:model} we briefly describe our model while in
\S\ref{sec:results} we present out results. Finally, in
\S\ref{sec:conc} we give our conclusions.

\section{Model}
\label{sec:model}

We use the methods described by \scite{benson02a} to evolve the
thermal and ionization properties of gas in the IGM and refer the
reader to that paper for a detailed discussion of the
calculations. Briefly, we solve the equations governing the evolution
of the ionization states and temperature of gas at a representative
range of density contrasts, beginning from shortly after the epoch of
recombination. The distribution of gas densities is drawn from the
distribution described by \scite{benson02a} which is chosen to
reproduce a given, time-dependent clumping factor (where clumping
factor is defined as $f_{\rm clump}=\langle \rho^2\rangle / \langle
\rho\rangle^2$ where $\rho$ is gas density in the IGM). We solve the
ionization and thermal evolution for densities spanning the range from
very underdense voids to densities comparable to those found in dark
matter halos. We can integrate over the suitably weighted gas
properties as a function of density to compute volume or mass weighted
quantities (such as the mean temperature of the IGM for example).

Both collisional and photoionization are considered in computing
ionization rates. Heating of the gas occurs through photoheating,
while cooling occurs due to atomic processes and Compton cooling off
CMB photons. Contributions to photoheating and photoionization from
both galaxies and quasars are included. We use the semi-analytic model
of \scite{benson02a} to compute the ionizing emissivity of galaxies as
a function of time, while for quasars we use the fitting function of
\scite{mhr99}. Note that the emissivity of galaxies will be affected
by preheating as we will describe below, but the quasar contribution is
fixed, since it is determined from observational measurements. For
galaxies, we assume that a fraction $f_{\rm esc}$ of all ionizing
photons produced are able to escape into the IGM and so contribute to
ionization and heating. Unless otherwise noted we will adopt $f_{\rm
esc}=0.1$ throughout this work \cite{leitherer,steidel}.

We adopt cosmological parameters $(\Omega_0,\Lambda_0,\Omega_{\rm
b},\sigma_8,h)=(0.3,0.7,0.045,0.93,0.7)$ consistent with current
observational constraints
(e.g. \pcite{cmb,hstkey,smith02,burles02}). \scite{benson02a}
considered photoionization by stars and quasars as the only energy
input into the IGM. \scite{madau01} showed that the IGM could be
heated to a higher adiabat by pregalactic outflows at high redshift.
To explore the effects of this preheating, we include a rapid
deposition of energy into the IGM at early times, in addition to the
photoionization. We characterize the energy input due to preheating by
the energy per baryon, $E_{\rm preheat}$. This energy is deposited in
the IGM at redshift $z_{\rm preheat}$. To be precise, the energy is in
fact added gradually over a short time centred on this redshift. This
allows for an easier numerical solution of the equations governing the
thermal and ionization state of the IGM. Such rapid preheating may be
relevant if the sources of the energy are Pop~III stars which
experience a strong negative feedback and so form a short-lived
population. We find that typically we can in fact increase the length
of the period over which energy is added significantly without
affecting our results. For example, in a model with $E_{\rm
preheat}=0.1$keV and $z_{\rm preheat}=9$ adding the energy over a
redshift interval of $\Delta z=3$ has no significant effect on our
results for galaxy luminosity functions (increasing $\Delta z$ to
approximately $z$ however results in the effects of preheating being
almost entirely removed). Our results are therefore equally valid for
both very rapid energy deposition and deposition occurring over a
significant fraction of a Hubble time.

We examine a homogeneous energy deposition since the filling factor of
pregalactic outflows is expected to be large \protect\cite{madau01}.
Recent numerical simulations have shown that outflows from
starbursting dwarf galaxies can enrich $\sim 20\%$ of the simulation
volume at the end of the cosmic dark ages (Thacker, Scannapieco \&
Davis 2002), while semi-analytical models that include H$_2$ cooling
in minihalos and the formation of `Population III' very massive stars
can yield filling factor of unity (Furlanetto \& Loeb 2003).

As described by \scite{benson02a} we use the resulting thermal history
of the IGM to compute the filtering mass \cite{gnedin00}, which in
turn allows us to determine the effects of the IGM temperature on the
accretion of gas into dark matter halos. This is input into the {\sc
galform} semi-analytic model of galaxy formation \cite{cole00} in
order to compute the luminosity function of galaxies. The {\sc
galform} model follows the formation of galaxies in a merging
hierarchy of dark matter halos. By calculating the rate at which gas
is able to cool into a star forming phase (and adopting simple rules
for the rate of star formation in that phase) {\sc galform} is able to
estimate the luminosity of galaxies as a function of time. The most
massive galaxies are typically built up through merging of smaller
systems (a process driven by dynamical friction). By simulating galaxy
formation in dark matter halos spanning a broad range of masses we are
able to construct the expected luminosity function of galaxies at the
present day.

The filtering mass is conventionally computed using the volume
averaged temperature of the IGM. However, unlike purely photo-heated
models the temperature distribution in a preheated IGM at late times
can be highly inhomogeneous (as will be discussed in
\S\ref{sec:therm}). As such, we consider a possibly more reasonable
approach in our preheated models, and compute filtering masses using
the density-dependent temperature predicted by our IGM model, for
several representative densities. Using the same probability distribution
function (PDF) for the density
distribution as used in our IGM model (see \pcite{benson02a}) we also
compute the fraction of the IGM's mass which exists in each of these
density bins. We then apply our galaxy formation model to compute the
properties of galaxies existing in dark matter halos. For each such
halo modeled we select one of the density-dependent filtering
masses. This selection is done at random, weighting by the mass
fraction present in each density bin such that the probability for a
halo to exist in each density range is proportional to the IGM mass in
that density range.\footnote{Here, we are treating the density in our
IGM model as describing the large scale density environment within
which a dark matter halo forms (note that the density PDF used is
equally applicable to dark matter and gas when we are considering
large scales). While this seems to be the most reasonable approach
(since the filtering mass prescription is based upon a linear theory
calculation), it has not been tested in numerical simulations and so
must be viewed with some degree of caution. It should also be noted
that weighting the selection of filtering masses by the mass fraction
in each density bin assumes that dark matter halo formation is
unaffected by the large scale density environment.}

Unlike \scite{benson02a} we do not allow the ionizing background to
heat gas already in halos (due to the high computational cost of this
calculation). As shown by Benson et al. this causes only a minor
additional suppression of galaxy formation. Its effect will be even
more negligible in this work, where we consider filtering masses which
are much higher than those in Benson et al. We adopt the same
parameters for the semi-analytic model as did \scite{benson02a}, with
the exception of using the more realistic value $\Omega_{\rm b}=0.045$
for the baryon density parameter. As we are interested in whether
preheating can produce a galaxy luminosity function with a flat
faint-end slope (as is observed), we switch off the effects of
supernova feedback in {\sc galform}. This local heating mechanism is
normally required to produce a flat luminosity function at the present
epoch.
  
\section{Results}
\label{sec:results}

Theoretical modeling of the first stars and galaxies provides a
valuable guide for the range of preheating energies and redshifts
which should be considered. \scite{lowenstein01} suggests that Pop~III
stars may preheat the intracluster medium at a level of $\sim 0.1$keV
per baryon at $z\gsim 10$. \scite{madau01} find preheating energies $<
0.1$keV at $z\approx 9$ from pregalactic winds. In order to determine
what constraints galaxy formation can place on preheating scenarios we
choose to study models spanning the range $E_{\rm
preheat}=0.05$--$0.3$keV and $z_{\rm preheat}=6$--$12$. This
incorporates the theoretical estimates described above, and also
covers more extreme models, both higher in energy (motivated by
observations of X-ray clusters) and with more recent preheating (when
preheating occurs recently there is less time for the IGM to cool and
return to its previous thermal state).

Note that all our models satisfy the limit on the thermal pressure of
intergalactic gas imposed by the lack of a Compton $y$-distortion to
the spectrum of the CMB observed by the {\it Cosmic Background
Explorer} ({\it COBE}) satellite. The mean thermal energy density
introduced into the IGM at $z_{\rm preheat}$ is
\begin{eqnarray}
U_{\rm IGM}&=&E_{\rm preheat} n_b(z_{\rm preheat}) =
3.6\times 10^{-13}\,{\rm ergs~cm^{-3}} \nonumber \\
 & & \times \left({E_{\rm preheat}\over {\rm keV}}
\right)\,\left({\Omega_b h^2\over 0.02}\right)\,\left({1+z_{\rm preheat}
\over 10}\right)^3.
\end{eqnarray}
Since $Ht_{\rm comp}\propto (1+z)^{-5/2}$ [assuming $H\propto
(1+z)^{3/2}$ as is appropriate for high redshifts], where $H$ is the
Hubble constant and $t_{\rm comp}$ is the Compton cooling time of hot
electrons off CMB photons,
\begin{equation}
t_{\rm comp}={3m_e {\rm c}\over 4 \sigma_T U_{\rm CMB}}=7.4\times 10^{15}~{\rm s}\;
\left({1+z\over 10}\right)^{-4},
\end{equation}
inverse Compton scattering will transfer all the energy released to
the CMB for $z>z_{\rm comp}=7h^{2/5}-1\approx 5$. Here $m_e$ is the
electron mass, $\sigma_T$ the Thomson cross-section, $U_{\rm CMB}$ is
the energy density of the CMB, and we have assumed a pure hydrogen
plasma such that the total number density of particles, $n_{\rm tot}$
is twice the number density of electrons, $n_{\rm e}$.  The amount of
$y$-distortion expected to the spectrum of the CMB is
\begin{eqnarray}
y&=&\left({U_{\rm IGM}\over 4 U_{\rm CMB}}\right)_{z=z_{\rm preheat}}=
2.16\times 10^{-5}\,\left({1+z_{\rm preheat}\over 10}\right)^{-1} \nonumber \\
 & & \times \left({\Omega_b h^2\over 0.02}\right) \left({E_{\rm preheat}\over {\rm keV}}\right)
\end{eqnarray} 
(Sunyaev \& Zel'dovich 1980). The {\it COBE} satellite measured
$y<1.5\times 10^{-5}$ at the 2$\sigma$ level (Fixsen et al. 1996),
implying $E_{\rm preheat}<0.07\,(1+z_{\rm preheat})\,$ keV. This limit
holds for $z_{\rm preheat}>z_{\rm comp}$; at redshifts $z_{\rm
preheat}<z_{\rm comp}$ the cooling time for Comptonization exceeds the
expansion timescale, and only a small fraction of the thermal energy
released is transferred to the CMB.

\subsection{Sources of Preheating}

It is interesting at this stage to set some general constraints on the
early star-formation episode and stellar populations that may be
responsible for preheating the IGM at the levels envisaged here. Let
$\Omega_*$ be the mass density of stars formed at $z_{\rm preheat}$ in
units of the critical density, $E_{\rm SN}$ the mechanical energy
injected per SN event, and $f_{\rm w}$ the fraction of that energy
that is eventually deposited into the IGM. Denoting with $\eta$ the
number of SN explosions per mass of stars formed, one can write
\begin{equation}
{\Omega_* \over \Omega_b}={E_{\rm preheat}\over f_{\rm w}\eta E_{\rm SN} 
m_p}, 
\label{starb}        
\end{equation}
where $m_p$ is the proton mass. For a Salpeter initial mass function
(IMF) between 0.1 and 100 M$_\odot$, the number of Type II SN
explosions per mass of stars formed is $\eta=0.0074$ M$_\odot^{-1}$,
assuming all stars above 8 M$_\odot$ result in SNe II. Numerical
simulations of the dynamics of SN-driven bubbles from subgalactic
halos have shown that up to 40\% of the available SN mechanical
luminosity can be converted into kinetic energy of the blown away
material, $f_{\rm w}\approx 0.4$, the remainder being radiated away
(Mori et al. 2002). With $E_{\rm SN}=1.2\times 10^{51}\,$ergs,
equation (\ref{starb}) implies
\begin{equation}
\left({\Omega_*\over \Omega_b}\right)_{\rm sp}=0.05~(E_{\rm preheat}/0.1~{\rm 
keV}).
\end{equation}
SN-driven pregalactic outflows efficiently carry metals into
intergalactic space (Madau et al. 2001). For a normal IMF, the total
amount of metals expelled in winds and final ejecta (in SNe or
planetary nebulae) is about 1\% of the input mass. Assuming a large
fraction, $f_{\rm Z}=0.5$, of the metal-rich SN ejecta escape the
shallow potential wells of subgalactic systems, the star-formation
episode responsible for early preheating will enrich the IGM to a mean
level
\begin{equation}
\langle Z\rangle_{\rm sp}={0.01\,\Omega_*\,f_{\rm Z}\over \Omega_b}=
0.014\,Z_\odot~(E_{\rm preheat}/0.1~{\rm keV}),
\end{equation}
where we take $Z_\odot=0.02$. The weak C IV absorption lines observed
in the \Lya forest at $z=3-3.5$ imply a minimum universal metallicity
relative to solar in the range [C/H]$=-3.2$ to $-2.5$ (Songaila
1997). The metal abundances of the \Lya clouds may underestimate the
average metallicity of the IGM if there existed a significant warm-hot
gas phase component with a higher level of enrichment, as detected for
example in O VI (Simcoe, Sargent \& Rauch 2002).  Today, the
metallicity of the IGM may be closer to $\sim 1/3$ of Solar if the
metal productivity of galaxies within clusters is to be taken as
representative of the universe as a whole (e.g. Renzini
1997).\footnote{Note that a metallicity $\sim 0.5 Z_\odot$ at $z=10$
increases the gas radiative cooling rate to a level comparable to
inverse Compton cooling. Our calculations assume a cooling function
for a primordial plasma. At the low metallicities typical of \Lya
forest clouds, the thermal behaviour can be modeled to a good
approximation by a gas with primordial abundances (e.g. Sutherland \&
Dopita 1993).}\ Preheating energies in excess of $0.1\,$keV appear to
require values of $\Omega_*$ and $\langle Z\rangle$ that are
comparable to the total mass fraction in stars seen today
(e.g. Glazebrook et al. 2003) and well in excess of the minimum
enrichment of the IGM inferred at intermediate redshifts,
respectively.

\begin{figure*}
\begin{tabular}{cc}
\psfig{file=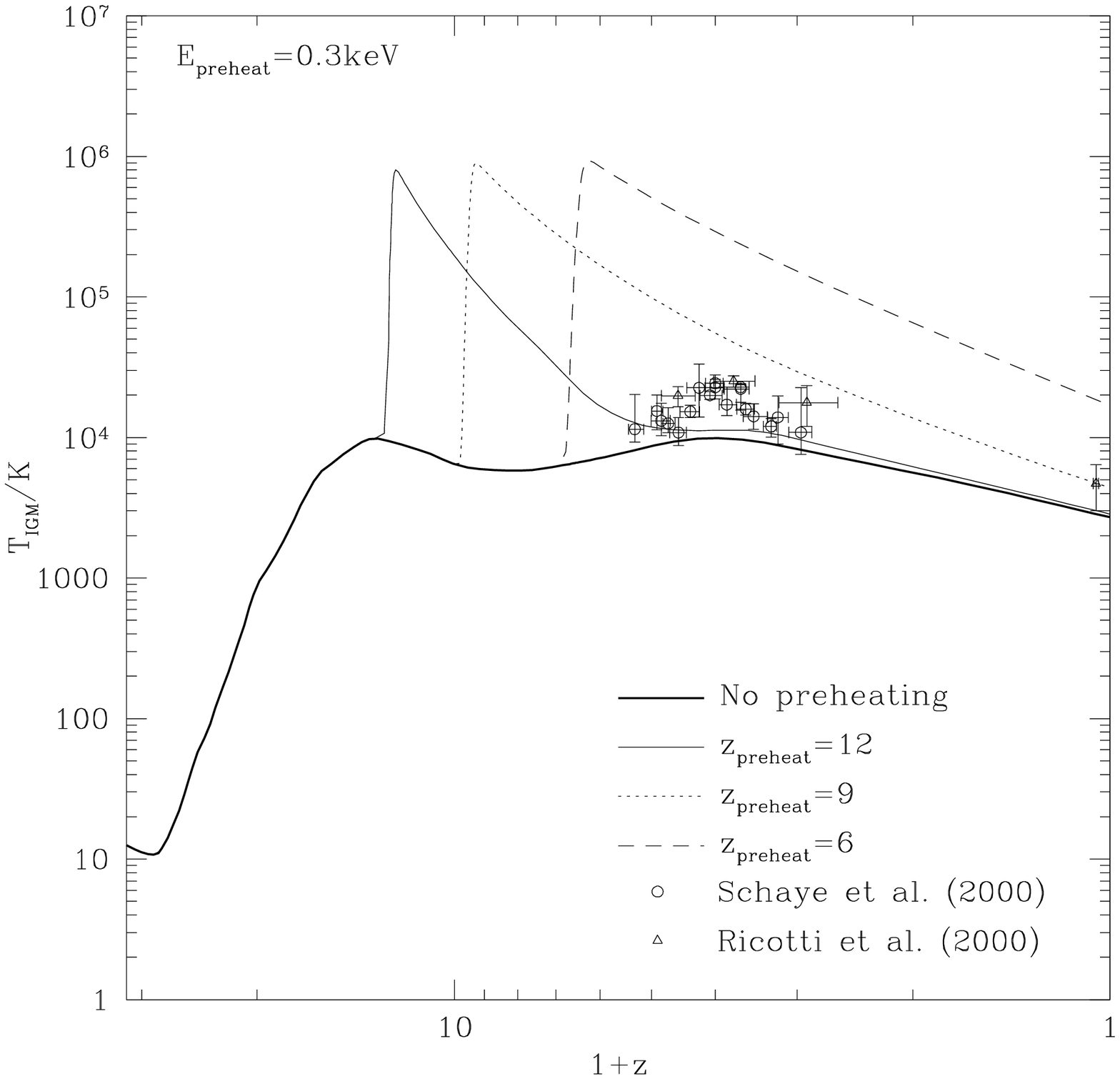,width=80mm} & \psfig{file=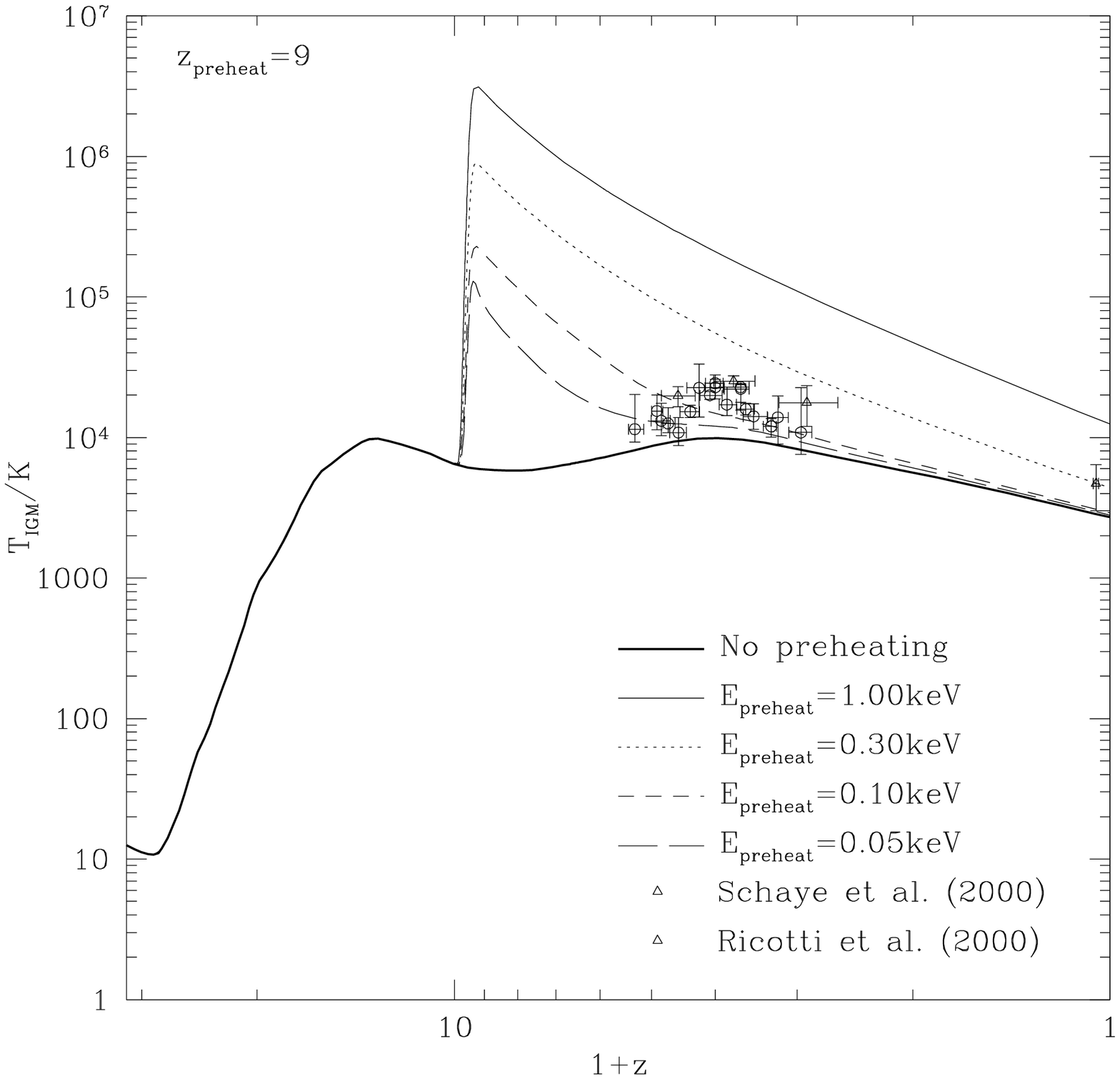,width=80mm}
\end{tabular}
\caption{The temperature of the IGM gas at mean density as a function
of redshift. The heavy, solid line shows the results for no
preheating. Points show the determinations of
\protect\scite{schaye2000} (circles) and \protect\scite{ricotti00}
(triangles) based on observations of quasar absorption
lines. \emph{Left-hand panel:} Thin lines show model results for
$E_{\rm preheat}=0.3$keV, and for three different values of $z_{\rm
reheat}$ as indicated in the figure. \emph{Right-hand panel:} Thin
lines show model results for $z_{\rm preheat}=9$, and for four
different values of $E_{\rm preheat}$ as indicated in the figure.}
\label{fig:temp}
\end{figure*}

Uncertainties in the early IMF make other preheating scenarios
possible and perhaps even more likely.  The very first generation of
metal-free Population III stars may have formed with an IMF biased
towards very massive members (i.e. stars a few hundred times more
massive than the Sun), quite different from the present-day Galactic
case (Bromm, Coppi \& Larson 1999; Abel, Bryan \& Norman
2000). Population III stars with main-sequence masses of approximately
$140-260\,$ M$_\odot$ will encounter the electron-positron pair
instability and be completely disrupted by a giant nuclear-powered
explosion (Heger \& Woosley 2002). A fiducial 200 M$_\odot$ Population
III star will explode with a kinetic energy at infinity of $E_{\rm
SN}=4\times 10^{52}\,$ ergs, injecting about 90 M$_\odot$ of metals
(Heger \& Woosley 2002). For a very `top-heavy' IMF with $\eta=0.005$
M$_\odot^{-1}$, equation (\ref{starb}) now yields (assuming $f_{\rm
w}=1$)
\begin{equation}
\left({\Omega_* \over \Omega_b}\right)_{\rm III}=
0.001~(E_{\rm preheat}/0.1\,{\rm keV}),
\end{equation}
and a mean IGM metallicity (assuming $f_{\rm Z}=1$)
\begin{equation}
\langle Z\rangle_{\rm III}={0.45\,\Omega_*\,f_{\rm Z}\over \Omega_b}=
0.02\,Z_\odot~(E_{\rm preheat}/0.1\,{\rm keV}).
\end{equation}
This scenario can yield large preheating energies by converting only a
small fraction of the comic baryons into Population III stars.
This is even more true for preheating from winds produced by an early, 
numerous population of faint miniquasars.
\footnote{Because the number density of {\it bright} quasi-stellar
objects (QSOs) at $z>3$ is low (Fan et al. 2001), the thermal and
kinetic energy they expel into intergalactic space must be very large
to have a global effect, i.e. for their blastwaves to fill and preheat
the universe as a whole. The energy density needed for rare, luminous
QSOs to shock-heat the entire IGM would in this case violate the {\it
COBE} limit on $y$-distortion (Voit 1994, 1996).}~Thin disk accretion
onto a Schwarzchild black hole releases about 50 MeV per baryon. If a
fraction $f_{\rm w}$ of this energy is used to drive an outflow and is
ultimately deposited into the IGM, the accretion of a trace amount of
the total baryonic mass onto early black holes,
\begin{equation}
{\Omega_{\rm BH}\over \Omega_b}={E_{\rm preheat}\over f_{\rm w}\,50\,{\rm MeV}}
=2\times 10^{-6}~f_{\rm w}^{-1}\,(E_{\rm preheat}/0.1\,{\rm keV}),
\end{equation}
may then suffice to preheat the whole universe. Note that this value
is about $50\,f_{\rm w}$ times smaller than the density parameter of the
supermassive variety found today in the nuclei of most nearby
galaxies, $\Omega_{\rm SMBH}\approx 2\times 10^{-6}\,h^{-1}$ (Merritt
\& Ferrarese 2001).

\subsection{Thermal Evolution}
\label{sec:therm}

In Figure~\ref{fig:temp} we show the thermal history of IGM gas at the
mean density of the Universe for a variety of $E_{\rm preheat}$ and
$z_{\rm preheat}$. In each case, the gas is initially heated by
photoionization from the first stars (beginning at $z\approx 30$). The
preheating energy causes a rapid increase in the temperature at
$z_{\rm preheat}$. Note that the temperature never becomes as high as
$2 E_{\rm preheat}/3{\rm k_B}$ since the heating ionizes the gas,
freeing electrons and thereby increasing the number density of
particles (recall that $E_{\rm preheat}$ specifies the energy per {\it
baryon}). We include the effects of inverse Compton cooling, adiabatic
expansion, and atomic cooling. For gas close to the mean density of
the Universe, Compton cooling and adiabatic expansion dominate the
cooling of gas after preheating. However, for high densities
(i.e. densities typical of regions forming galaxies) cooling is
dominated by atomic processes. Consequently, at high densities the gas
is typically cooler than the results shown in Fig.~\ref{fig:temp}
(which are for gas at mean density). This will have important
consequences for the filtering mass and galaxy luminosity function as
will be discussed in \S\ref{sec:filtering} and \S\ref{sec:LF}, where
we will compute filtering masses as a function of density, and follow
galaxy formation for each different filtering mass. A model with no
preheating is also shown, whose only heat source is therefore
photoheating.

We compare our model results to the observational determinations of
\scite{schaye2000} and \scite{ricotti00}. This comparison will be used
to discard models which are strongly inconsistent with the data. It is
clear that the measurements of the IGM temperature at $z\sim 3$ rule
out models in which $E_{\rm preheat}$ is too high, or $z_{\rm
preheat}$ is too low. For sufficiently low $E_{\rm preheat}$ or high
$z_{\rm preheat}$ the IGM is able to recover to close to the thermal
state of the no preheating case, which lies close to the data, by
$z=3$. The result is that models which heat to $T_{\rm IGM}\gsim
10^6\,$K at $z\lsim 10$ are inconsistent with the $z\approx 3$
temperature data. The models which adequately fit the temperature data
are indicated in Table~\ref{tb:models}, where we also indicate which
models are consistent with the measured Compton $y$-distortion in the
CMB. We consider only those models consistent with both constraints
for the remainder of the paper. (Note that we will typically not plot
lines for the $E_{\rm preheat}=0.05$keV models to avoid overcrowding
the figures.) Table~\ref{tb:models} also lists the optical depth to
Thomson scattering for CMB photons, $\tau_{\rm e}$. There is very
little variation between the models since the bulk of hydrogen
reionization occurs through photoionization prior to preheating. Our
models are consistent with the constraints on the optical depth from
the {\it WMAP} experiment \cite{kogut03}, $\tau=0.17\pm 0.04$, only at
the $1.5\sigma$ level. This discrepancy can be resolved somewhat by
exploring models with higher values of $f_{\rm esc}$. In
Table~\ref{tb:models} we show results for models with $E_{\rm
preheat}=0.1$keV and $z_{\rm preheat}=9$ and $12$, for $f_{\rm
esc}=20\%$ and $50\%$. All are consistent with the measured IGM
temperature and the Compton-$y$ constraint, but produce higher optical
depths due to partial photoionization of hydrogen at high
redshifts. The models with $f_{\rm esc}=50\%$ achieve $\tau=0.15$,
very close to the WMAP value.

\begin{table*}
\caption{Properties of the nine $f_{\rm esc}=10\%$ models and
additional higher $f_{\rm esc}$ models are considered. Columns 1 and 2
list the preheating energy and redshift respectively. Column 4 notes
whether the model produces a reasonable match to the measured IGM
temperature at $z\sim 3$, Column~5 indicates whether the model is
consistent with the measured limit on the Compton $y$-distortion of
the CMB, while Column~6 lists the optical depth to Thomson scattering
for CMB photons.}
\label{tb:models}
\begin{center}
\begin{tabular}{cccccc}
\hline
$E_{\rm preheat}$/keV & $z_{\rm preheat}$ & $f_{\rm esc}$ & Fits $T_{\rm IGM}$? & $y$-distortion OK? & $\tau_{\rm e}$ \\
\hline
0.00 & --- & 10\% & $\surd$ & $\surd$ & 0.11 \\
0.05 & 6 & 10\% & $\frac{1}{2}\surd$ & $\surd$ & 0.11 \\
0.05 & 9 & 10\% & $\surd$ & $\surd$ & 0.11 \\
0.05 & 12 & 10\% & $\surd$ & $\surd$ & 0.11 \\
0.10 & 6 & 10\% & $\times$ & $\surd$ & 0.11 \\
0.10 & 9 & 10\% & $\surd$ & $\surd$ & 0.11 \\
0.10 & 12 & 10\% & $\surd$ & $\surd$ & 0.11 \\
0.30 & 6 & 10\% & $\times$ & $\surd$ & 0.11 \\
0.30 & 9 & 10\% & $\times$ & $\surd$ & 0.11 \\
0.30 & 12 & 10\% & $\surd$ & $\surd$ & 0.12  \\
1.00 & 6 & 10\% & $\times$ & $\times$ & 0.11 \\
1.00 & 9 & 10\% & $\times$ & $\times$ & 0.11 \\
1.00 & 12 & 10\% & $\surd$ & $\times$ & 0.12 \\
\hline
0.10 & 9 & 20\% & $\surd$ & $\surd$ & 0.13 \\
0.10 & 12 & 20\% & $\surd$ & $\surd$ & 0.13 \\
0.10 & 9 & 50\% & $\surd$ & $\surd$ & 0.15 \\
0.10 & 12 & 50\% & $\surd$ & $\surd$ & 0.15 \\
\hline
\end{tabular}
\end{center}
\end{table*}

Before considering the effect of preheating on the galaxy luminosity
function we examine two other predictions from our model---the
ionization state of the IGM and the entropy of IGM gas.

\subsection{Ionization States}
\label{sec:ion}

\begin{figure}
\psfig{file=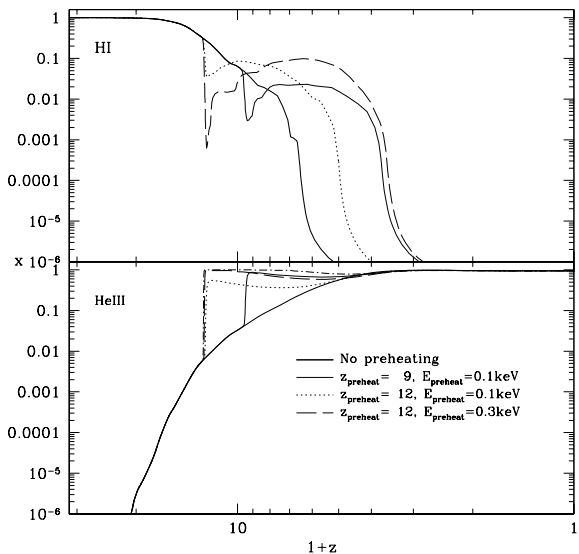,width=80mm}
\caption{The fractional density of \HI\ (i.e. $n_{{\rm H}{\sc
i}}/n_{\rm H}$; upper panel), and that of \HeIII\ (i.e. $n_{{\rm
He}{\sc iii}}/n_{\rm He}$; lower panel), as a function of redshift for
our models. Results are shown for all models which adequately fit the
$z\approx 3$ temperature data and the Compton $y$-distortion
constraint (Table~\protect\ref{tb:models}), for an assumed escape
fraction $f_{\rm esc}=$10\%.}
\label{fig:ion}
\end{figure}

The large amount of energy deposited into the IGM during preheating
will necessarily affect the ionization state of gas in the IGM. The
fractional densities of \HI\ and \HeIII\ in our models are shown in
Figure~\ref{fig:ion}. \HI\ and \HeII\ are collisionally ionized at
$z_{\rm preheat}$ in all of our models.

In some cases, the neutral fraction is of order $0.001$ or higher. As
such, these models would still produce a Gunn-Peterson trough after
$z_{\rm preheat}$. \HI\ is replenished after $z_{\rm preheat}$ by
radiative recombinations (the features in the curves are due to
changes in recombination rates as the IGM cools), and is finally
almost fully ionized between $z\approx 6$ and $z\approx 2$ through
photoionizations. The neutral fraction at these redshifts is often
larger than in the no preheating case. This occurs because preheating
exerts a strong negative feedback on galaxy formation, resulting in
fewer ionizing photons being available at these redshifts and
consequently a higher neutral fraction. This may allow these models to
explain the Gunn-Peterson troughs seen in the spectra of the most
distant Sloan Digital Sky Survey quasars \cite{becker01} at $z\approx
6$. However, the observed lack of a Gunn-Peterson trough at $z\lsim 6$
clearly rules out these models. A possible solution to this problem
lies in increasing the escape fraction of ionizing photons as
discussed below.

For helium (lower panel), we see that preheating typically causes
ionization to \HeIII, which then remains at an almost constant level
until $z=0$. Note that this is at variance with more conventional
scenarios in which the double reionization of helium occurred later,
at a redshift of 3 or so (see Kriss et al. 2001, and references
therein), due to the integrated radiation emitted above 4~Ryd by QSOs
(but see Oh et al. 2001).

\begin{figure}
\psfig{file=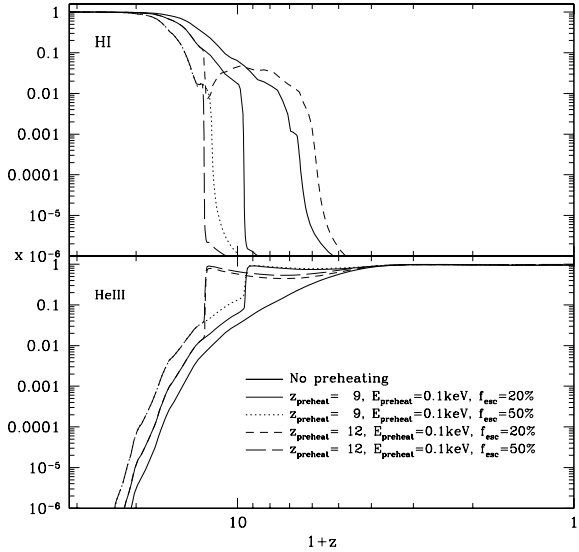,width=80mm}
\caption{The fractional density of \HI\ (i.e. $n_{{\rm H}{\sc
i}}/n_{\rm H}$; upper panel), and that of \HeIII\ (i.e. $n_{{\rm
He}{\sc iii}}/n_{\rm He}$; lower panel), as a function of redshift for
our models with higher $f_{\rm esc}$.}
\label{fig:fescion}
\end{figure}

As noted above, the presence of a significant fraction of neutral
hydrogen at $z\lsim 6$ in our preheated models would conflict with the
lack of an observed Gunn-Peterson effect at these redshifts. A
possible solution to this problem is to increase the rate of
photoionization by increasing the escape fraction, $f_{\rm esc}$ for
galaxies. Figure~\ref{fig:fescion} shows the ionization fractions for
models with $E_{\rm preheat}=0.1$keV, $z_{\rm preheat}=9$ and $12$ and
with increased escape fractions of $f_{\rm esc}=20$\% and 50\%. While
there is no observational evidence for such high escape fractions at
low-redshifts our ignorance of the nature of very high redshift
galaxies makes it interesting to consider the consequences of such
high escape fractions.

Increasing $f_{\rm esc}$ to 20\% is sufficient to reduce the neutral
fraction to negligible levels in the $z_{\rm preheat}=9$ model, while
a higher escape fraction still is required for the $z_{\rm
preheat}=12$ model. We conclude that an increased escape fraction will
remove the residual neutral hydrogen which is problematic for our
$f_{\rm esc}=10\%$ models. Furthermore, this higher escape fraction
has only a small impact on the thermal evolution of the IGM, and the
filtering mass and luminosity functions remain largely unchanged. Our
conclusions regarding these quantities in the remainder of the paper
are therefore equally valid for these higher escape fractions. It is
also interesting to note that these higher escape fractions result in
somewhat better agreement with the {\it WMAP} optical depth
measurements. An escape fraction of 20\% results in $\tau=0.13$ while
$f_{\rm esc}=50\%$ results in $\tau=0.15$.

\begin{figure}
\psfig{file=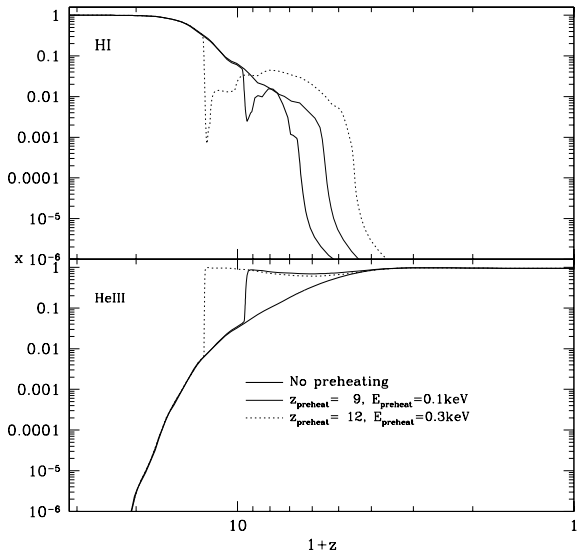,width=80mm}
\caption{The fractional density of \HI\ (i.e. $n_{{\rm H}{\sc
i}}/n_{\rm H}$; upper panel), and that of \HeIII\ (i.e. $n_{{\rm
He}{\sc iii}}/n_{\rm He}$; lower panel), as a function of redshift for
models computed using density-dependent filtering masses.}
\label{fig:localion}
\end{figure}

As described in \S\ref{sec:model} we have also performed calculations
using a density-dependent filtering mass in order to approximately
account for the significant inhomogeneity in the IGM temperature in
preheated models. Figure~\ref{fig:localion} shows the ionization
fractions for two such models. Galaxy formation in high density
regions is significantly less suppressed in these models since, as we
will see in \S\ref{sec:filtering}, the temperature and filtering mass
are lower. As such, the neutral hydrogen fractions in these models
drop to very low values, albeit somewhat later than a model with no
preheating. This helps reconcile these models with the SDSS quasar
observations, although clearly some additional increase in $f_{\rm
esc}$ is still required.

\subsection{Entropy}
\label{sec:entropy}

\begin{figure}
\psfig{file=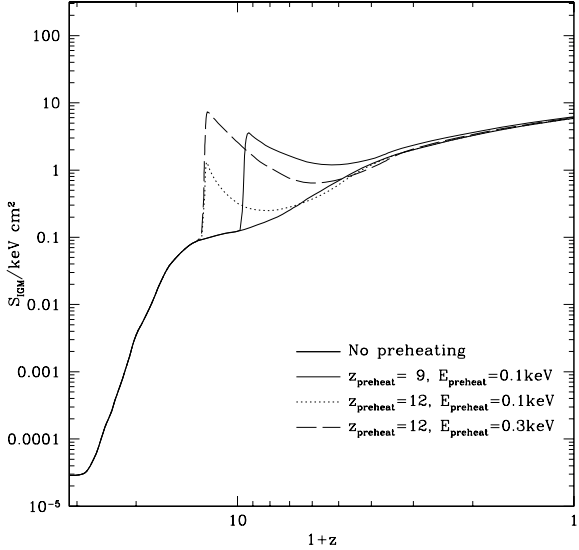,width=80mm}
\caption{The entropy, $S={\rm k_B}T/n_{\rm e}^{2/3}$, of IGM gas in
our models shown as a function of redshift. Results are shown for all
models which adequately fit the $z\approx 3$ temperature data and the
Compton $y$-distortion constraint (Table~\protect\ref{tb:models}).}
\label{fig:entropy}
\end{figure}

Preheating has been suggested as the origin of the entropy floor seen
in clusters of galaxies \cite{ponmanetal}. These observations imply an
``entropy'' of $S(={\rm k_B}T/n_{\rm e}^{2/3})\sim 100$keV cm$^2$ for
gas at $z=0$. We show, in Figure~\ref{fig:entropy}, the entropy of IGM
gas in our models as a function of redshift. The entropy is never
constant (as would be expected for gas cooling by adiabatic expansion)
due to the other cooling and heating processes included in our
calculation. Note that none of our models ever reach $S=100$keV
cm$^2$. The requirement that the IGM temperature match that which is
measured at $z\lsim 4$ limits the amount of entropy which can be
deposited into the IGM (the entropy produced by preheating is
increased by increasing $E_{\rm preheat}$ and/or decreasing $z_{\rm
preheat}$, both of which tend to result in temperatures which are too
high at $z\approx 3$), while \pcite{ohajb} note that preheating must
occur prior to $z\approx 2$ in order to affect the cores of clusters.
Alternatively, entropy generation spatially localized to regions which
are destined to become clusters could circumvent these
constraints. Finally, we note that the use of a density-dependent
filtering mass has little effect on the volume-averaged entropies
shown in Fig.~\ref{fig:entropy}. For very dense regions, little
entropy is generated by the preheating models considered here,
although considerable entropy \emph{is} produced through photo-heating
at late times reaching $S=50$keV cm$^2$ for regions with densities
comparable to dark matter halos at $z=0$.

\subsection{Filtering Mass}
\label{sec:filtering}

The filtering mass is central to our calculation of the effects of
preheating on the luminosity function of galaxies.  If the IGM has a
non-zero temperature, then pressure forces will prevent gravitational
collapse of the gas on small scales. For gas at constant temperature,
and ignoring the expansion of the Universe, the effects of pressure on
the growth of density fluctuations in the gas are described by a
simple Jeans criterion, such that density fluctuations on mass scales
below the Jeans mass $M_{\rm J}$ are stable against collapse. However,
this simple criterion needs to be modified in the case of an expanding
Universe in which the gas temperature is a function of
time. \scite{gnedinhui} have obtained an analytical description of the
effects of gas pressure in this case. From linear perturbation
analysis, the growth of density fluctuations in the gas is suppressed
for comoving wavenumbers $k>k_{\rm F}$, where the critical wavenumber
$k_{\rm F}$ is related to the Jeans wavenumber $k_{\rm J}$ by
\begin{equation}
{1 \over k^2_{\rm F}(t)} = {1 \over D(t)} \int_0^t \d t^\prime
a^2(t^\prime) {\ddot{D}(t^\prime) + 2 H(t^\prime) \dot{D}(t^\prime)
\over k^2_{\rm J}(t^\prime)} \int_{t^\prime}^t {\d t^{\prime\prime}
\over a^2(t^{\prime\prime})}
\end{equation}
and $k_{\rm J}$ is defined as
\begin{equation}
k_{\rm J} =  a \left(4\pi \G \bar{\rho}_{\rm tot} {3 \mu m_{\rm H} \over 5 k_B
\overline{T}_{\rm IGM}} \right)^{1/2}.
\label{eq:kF}
\end{equation}
In the above, $\bar{\rho}_{\rm tot}$ is the mean \emph{total} mass
density including dark matter, $D(t)$ and $H(t)$ are the linear growth
factor and Hubble constant respectively as functions of cosmic time
$t$, and a dot over a variable represents a derivative with respect to
$t$. \scite{gnedinhui} define $\overline{T}_{\rm IGM}$ to be the
volume-weighted mean temperature of the IGM. We therefore compute the
volume weighted temperature of IGM gas from our IGM model by averaging
over the temperatures of gas at each density considered in the
calculations. Such a global approach seems reasonable if the IGM
temperature is reasonably homogeneous, such as happens in the case of
a purely photoionized IGM (i.e. with no preheating). With preheating
however, there can be considerable inhomogeneity in the IGM
temperature since, after preheating, at the mean density cooling is
dominated by Compton cooling and adiabatic expansion (both of which
cool at a rate proportional to the gas density), while at high
densities atomic cooling processes dominate (which are proportional to
gas density squared). In this case it may be more realistic to compute
the filtering mass as a function of density, using the
density-dependent temperature in eqn.~(\ref{eq:kF}) to do so. We will
use the volume-weighted temperature of the IGM to compute filtering
masses unless stated otherwise, but consider the alternative approach
also.

\begin{figure}
\psfig{file=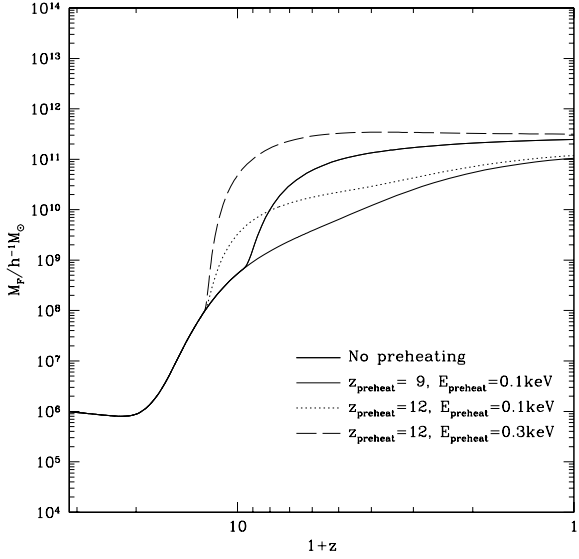,width=80mm}
\caption{The filtering mass as a function of redshift in our
models. Results are shown for all models which adequately fit the
$z\approx 3$ temperature data and the Compton $y$-distortion
constraint (Table~\protect\ref{tb:models}).}
\label{fig:filtering}
\end{figure}

The above expression for $k_{\rm F}$ accounts for arbitrary thermal
evolution of the IGM, through $k_{\rm J}(t)$. Corresponding to the
critical wavenumber $k_{\rm F}$ there is a critical mass $M_{\rm F}$
which we will hereafter call the filtering mass, defined as
\begin{equation}
M_{\rm F}=(4 \pi /3) \bar{\rho}_{\rm tot} (2 \pi a/k_{\rm F})^3
\label{eq:MF}
\end{equation}
The Jeans mass $M_{\rm J}$ is defined analogously in terms of $k_{\rm
J}$.  In the absence of pressure in the IGM, a halo of mass $M_{\rm
tot}$ would be expected to accrete a mass $(\Omega_{\rm
b}/\Omega_0)M_{\rm tot}$ in gas when it collapsed. \scite{gnedin00}
found that in cosmological gas-dynamical simulations with a
photoionized IGM, the average mass of gas $M_{\rm gas}$ which falls
into halos of mass $M_{\rm tot}$ can be fit with the formula
\begin{equation}
M_{\rm gas} =  {\left( \Omega_{\rm b} / \Omega_0 \right) M_{\rm tot} \over
[1 + (2^{1/3}-1)M_{\rm F}/M_{\rm tot}]^3}
\label{eq:Mgaccr}
\end{equation}
with the same value of $M_{\rm F}$ as given by equations~(\ref{eq:kF})
and (\ref{eq:MF}). The denominator in the above expression thus gives
the factor by which the accreted gas mass is reduced because of the
IGM pressure. Specifically, $M_{\rm F}$ gives the halo mass for which
the amount of gas accreted is reduced by a factor 2 compared to the
case of no IGM pressure. The resulting filtering mass for each of our
models is shown in Figure~\ref{fig:filtering}. The filtering mass
begins to rise at $z\approx 20$ due to the initial photoheating of the
IGM by early star formation. At the epoch of preheating the filtering
mass begins to rise sharply. For models with large $E_{\rm preheat}$,
the filtering mass subsequently remains almost constant to $z=0$.

It is important to note that the filtering mass prescription results
in a much more aggressive suppression of galaxy formation than the
simpler prescription in which halos with virial temperature $T_{\rm
vir}<T_{\rm IGM}$ are assumed to be unable to form galaxies. By $z=0$,
the hottest model we consider has $T_{\rm IGM}\approx 4000$K,
corresponding to the virial temperature of a $10^8h^{-1}M_\odot$ halo,
while the filtering mass for this model is a few times
$10^{11}h^{-1}M_\odot$. The ``thermal memory'' of the IGM as
encapsulated in the filtering mass is therefore of crucial importance
in determining the extent to which galaxy formation is
suppressed. Consequently, it would be extremely valuable to conduct
tests of the filtering mass prescription in preheated N-body
simulations of galaxy formation to validate its use in this regime.

For the no preheating case the filtering mass is approximately
$10^{11}h^{-1}M_\odot$ at $z=0$. Three of our models produce a
filtering mass at $z=0$ which is within a factor of three of this
value. The remaining two predict filtering masses roughly an order of
magnitude larger. In these latter two models, the filtering mass is
comparable to the mass of halos thought to host $L_*$ galaxies. As
such, we may expect these models to produce a dearth of $L_*$ and
fainter galaxies. Note that the filtering mass resulting from a given
$E_{\rm preheat}$ depends strongly on $z_{\rm preheat}$.

\begin{figure}
\psfig{file=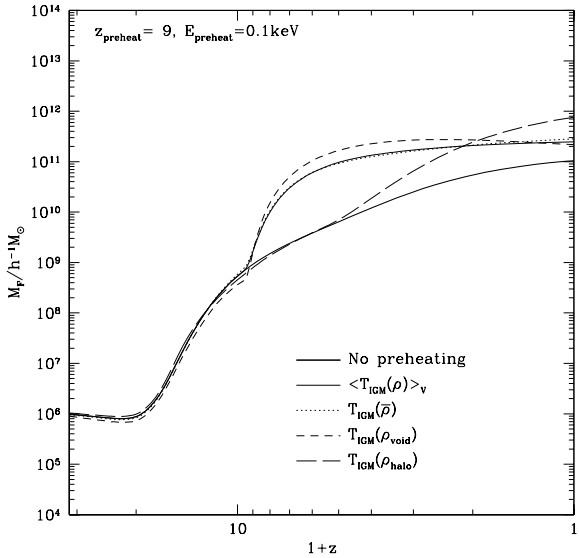,width=80mm}
\caption{The filtering mass as a function of redshift for $E_{\rm
preheat}=0.1$keV and $z_{\rm preheat}=9$. The thin solid line
indicates the result when the volume average IGM temperature is used
in eqn.~(\protect\ref{eq:kF}), while the dotted line shows the result
for gas at mean density. Short and long-dashed lines indicate the
filtering mass for gas with density comparable to that of a void and
of a dark matter halo at redshift zero respectively. For reference,
the heavy, solid line shows the result for no preheating.}
\label{fig:localfiltering}
\end{figure}

As discussed above, the IGM temperature can be highly inhomogeneous in
a preheated universe. As such, the use of a volume averaged IGM
temperature in eqn.~(\ref{eq:kF}) may be inappropriate. In
Fig.~\ref{fig:localfiltering} we show the filtering mass computed for
a model with $E_{\rm preheat}=0.1$keV and $z_{\rm preheat}=9$, using
the temperature history of gas at several different densities, and
compare this to the result obtained using the volume averaged
temperature. Not surprisingly, the filtering mass of gas at mean
density is very similar to that obtained using a volume averaged
temperature. Furthermore, very low density gas (which will form part
of a void at $z=0$) has a filtering mass very similar to that of gas
at mean density, since at low densities the dominant cooling
mechanisms (Compton cooling and adiabatic expansion) are proportional
to density. Fig.~\ref{fig:localfiltering} also shows the filtering
mass for gas which is at a density similar to that of a virialized
halo at $z=0$. Prior to $z\approx 2$, the filtering mass for this gas
is much lower than that for gas at mean density, lying close to the
filtering mass for a model with no preheating. Here, the high density
of the gas has allowed nearly all of the preheating energy to be
rapidly radiated away, and so it has little effect on the filtering
mass. After $z\approx 2$ the filtering mass for this high density gas
begins to rise due to photoheating (as this high density gas cools
abundance of neutral species increases, thereby raising the
photoheating rate), resulting in the filtering mass at $z=0$ being
somewhat higher than for gas at mean density. Nevertheless, over a
wide range of redshift the high density gas has a significantly lower
filtering mass than gas at mean density (and than that calculated
using a volume averaged IGM temperature). Since galaxies are expected
to form in high density regions this may have important consequences
for the galaxy luminosity function. We will explore this possibility
in \S\ref{sec:LF}.

\subsection{Luminosity Functions}
\label{sec:LF}

\begin{figure}
\psfig{file=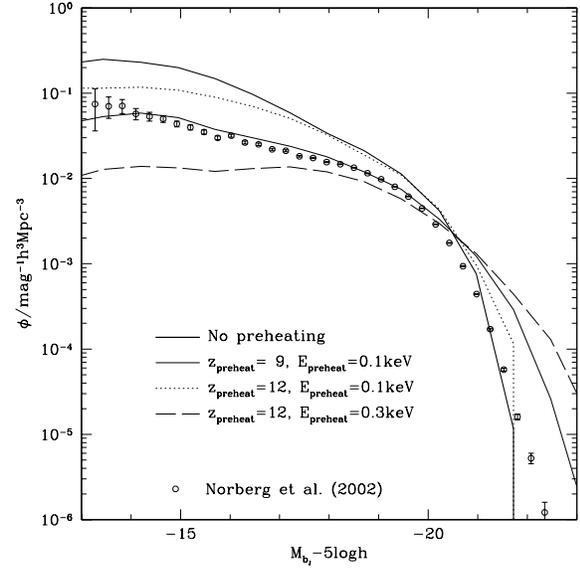,width=80mm}
\caption{B-band luminosity functions of galaxies at $z=0$, as
predicted by the semi-analytic model of \protect\scite{benson02a} and
using the filtering masses shown in Fig.~\protect\ref{fig:filtering},
are shown as lines. The observational determination of
\protect\scite{norberg02} is shown as circles.}
\label{fig:LFB}
\end{figure}

Figure~\ref{fig:LFB} shows the luminosity functions predicted by the
semi-analytic model of \scite{benson02a} when the filtering masses
from Figure~\ref{fig:filtering} are used. The model with no preheating
actually succeeds in matching the bright end of the luminosity function
rather well, but at the expense of over predicting the number of faint
galaxies by almost an order of magnitude. As was noted by
\scite{benson02a}, photoheating alone is not sufficient to explain the
paucity of faint galaxies.

As anticipated in \S\ref{sec:filtering} those models which produced a
filtering mass of several times $10^{11}h^{-1}M_\odot$ at $z=0$ result
in too few galaxies faintwards of $L_*$. As such, these models are
clearly inconsistent with the observational data (dashed line in
Figure~\ref{fig:LFB}). Of the remaining models we see that all perform
better at matching the faint end of the luminosity function than the
model with no preheating. In fact, the model with $E_{\rm
preheat}=0.1$keV and $z_{\rm preheat}=9$ produces a very good match
the the faint end of the luminosity function. However, this model
fails to produce a sufficiently sharp cut off at the bright end and so
over predicts the abundance of bright galaxies. This excess of bright
galaxies occurs because, when the faint end of the luminosity function
is sufficiently suppressed, too much gas remains available for cooling
at late times. This gas is then able to cool in massive halos,
producing an overabundance of bright galaxies. Note that the model
with $E_{\rm preheat}=0.1$keV and $z_{\rm preheat}=12$ does do
reasonably well at matching the bright end cut-off, but over predicts
at the faint end. Similarly, a model with $E_{\rm preheat}=0.05$keV
and $z_{\rm preheat}=9$ gets reasonably close to the bright end, but
again fails to suppress the faint end sufficiently. \scite{cole00}
were able to obtain a good match to the bright end of the luminosity
function in a model using SNe feedback to suppress the faint end. Our
failure to match the bright end in a preheated model is due to our use
of a higher $\Omega_{\rm b}$ than \scite{cole00} (0.045 instead of
0.02). With the higher $\Omega_{\rm b}$ used here (and which is now
preferred observationally) both preheated models and models with SNe
feedback suffer the same problems in trying to match the bright end of
the luminosity function (see \pcite{benson03b} for a detailed study of
the problem of the correctly matching the bright end of the luminosity
function).

\begin{figure}
\psfig{file=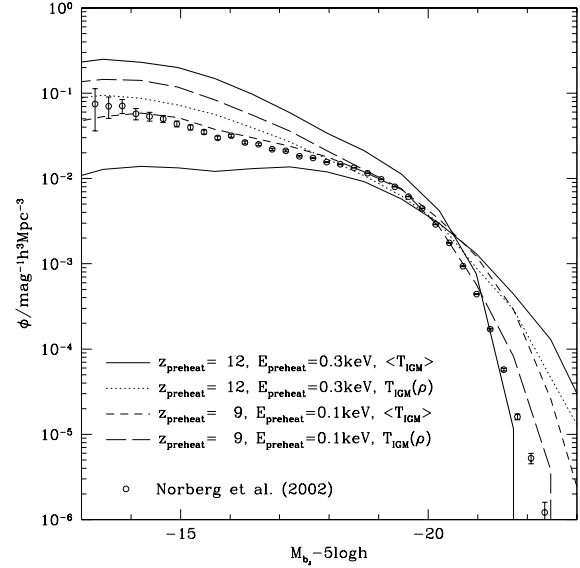,width=80mm}
\caption{B-band luminosity functions of galaxies at $z=0$, as
predicted by the semi-analytic model of \protect\scite{benson02a} and
using the filtering masses shown in
Fig.~\protect\ref{fig:localfiltering} (using the IGM temperature as a
function of density, $T_{\rm IGM}(\rho)$), are shown as lines, and are
compared to those computed using the filtering mass appropriate for a
volume averaged IGM temperature, $\langle T_{\rm IGM}\rangle$. The
observational determination of \protect\scite{norberg02} is shown as
circles.}
\label{fig:localLF}
\end{figure}

As discussed in \S\ref{sec:filtering}, if we compute the filtering
mass using the temperature history as a function of density in the IGM
(as opposed to using a volume averaged temperature history), we find
that high density regions have a much lower filtering mass than low or
average density regions. This will of course impact the luminosity
function of galaxies. In Fig.~\ref{fig:localLF} we show luminosity
functions computed using these alternative filtering masses.

We compare these luminosity functions to their counterparts computed
using the filtering mass for the volume averaged IGM temperature. The
luminosity functions are intermediate between that for the volume
averaged temperature calculation (since regions at average or lower
density produce luminosity functions of this type) and the no
preheating case (since high density regions produce luminosity
functions of this type). The $z_{\rm preheat}=9$, $E_{\rm
preheat}=0.1$keV model, which fits the observational data well using
the volume averaged filtering mass, is a poor match in the new
calculation. However, the $z_{\rm preheat}=12$, $E_{\rm
preheat}=0.3$keV model which previously caused too much suppression is
now a reasonable match to the faint-end of the luminosity
function. Clearly, stronger preheating is required to produce a good
match to the luminosity function when we account for the variation of
filtering mass with the environment.

\subsection{Local Group Satellites}

Finally, we examine the results of our models for very faint galaxies,
namely the satellite galaxies found in the Local Group. As first shown
by \scite{kwg93}, CDM models typically over predict the number of faint
satellite galaxies in the Local Group. Recently, \scite{benson02b}
examined the effect of photoheating on this abundance and concluded
that while photoheating produced a large (almost an order of
magnitude) reduction in the abundance of satellites, it was unable to
fully reconcile the theory and observations.

\begin{figure}
\psfig{file=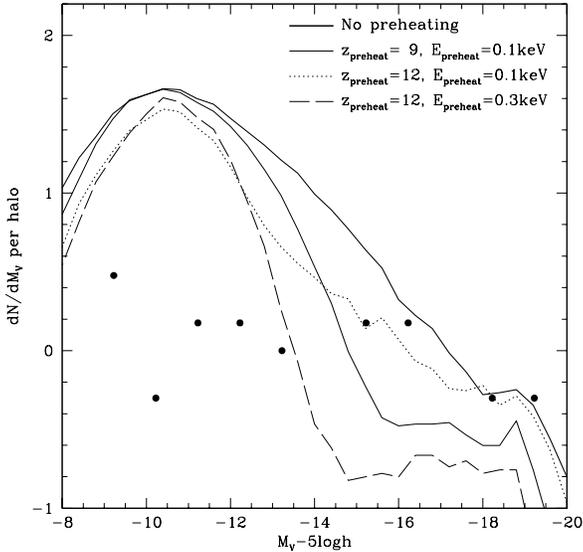,width=80mm}
\caption{The V-band luminosity function of satellite galaxies in the
Local Group. We plot the differential luminosity function per host
halo (n.b. we consider there to be two host halos---those of the Milky
Way and M31---in the Local Group). Points show the observational
result. Lines show the mean luminosity function of satellites in halos
of mass $10^{12}h^{-1}M_\odot$ in our models.}
\label{fig:satLF}
\end{figure}

In Figure~\ref{fig:satLF} we show as circles the V-band luminosity
function of Local Group satellite galaxies (from \pcite{benson02b})
per host halo (i.e. the halo in which the satellite orbits at the
present time). The results of our models are overlaid as lines. As
expected, the model with no preheating does not match the
observational data, and predicts too many faint
galaxies.\footnote{Note that this luminosity function differs from
that of \protect\scite{benson02b} since we do not include any feedback
due to supernovae here.} Adding in preheating causes a rapid increase
in the filtering mass just after $z_{\rm preheat}$. The effects of
this increase can be understood quite simply---in particular the
reader is referred to \scite{benson03a} who give a detailed discussion
of the effects of the filtering mass on the luminosity
function. Briefly, after $z_{\rm preheat}$, galaxy formation will be
suppressed in dark matter halos with mass less than the filtering
mass. Since the typical formation redshift of halos increases as the
halo mass decreases we expect the filtering mass to steepen the
luminosity function for faint galaxies. For halos with mass comparable
to the filtering mass we expect a flattening of the luminosity
function as the filtering mass gradually causes more suppression as
halo mass decreases.

This is exactly what is seen in Figure~\ref{fig:satLF}. For example,
the thin solid line ($E_{\rm preheat}=0.1$keV, $z_{\rm preheat}=9$) is
much steeper than the no preheating model (heavy solid line) in the
range $M_{\rm V}-5\log h=-12$ to $-15$, and then rapidly flattens in
the range $M_{\rm V}-5\log h=-16$ to $-18$. Evidently, none of these
models is able to satisfactorily fit the observational data. While the
flattening at bright magnitudes helps match the observed luminosity
function (e.g. the model with $E_{\rm preheat}=0.1$keV, $z_{\rm
preheat}=12$ does well brightwards of $M_{\rm V}-5\log h=-15$), the
steeper slope at faint magnitudes results in an overabundance of faint
satellites. Not surprisingly, no model using density-dependent
filtering masses performs any better at matching the Local Group
luminosity function.

\section{Conclusions}
\label{sec:conc}

We have calculated the thermal evolution of the IGM when it is rapidly
preheated at a given redshift. Observations of the temperature of the
IGM at $z\approx 3$ allow us to rule out models in which this
preheating occurs too late or to too high a temperature (simply
speaking, the IGM must have sufficient time before $z=3$ to cool down
after preheating). Unlike a purely photoionized IGM, the temperature
after preheating can become highly inhomogeneous, since different
cooling mechanisms dominate for gas of different
densities. Consequently, the effects of preheating on galaxy formation
depend strongly on whether we compute filtering masses using the
volume averaged IGM temperature, or a local, density-dependent
temperature.

Preheating causes an early reionization of the Universe, but in most
cases using a globally averaged filtering mass hydrogen is able to
mostly recombine before becoming highly ionized again at late times
through photoionization by stars and quasars. When density-dependent
filtering masses are used hydrogen does not recombine after
preheating, although full reionization is delayed relative to a model
with no preheating. For an escape fraction similar to current
observational limits we find that after preheating there is a
significant fraction of neutral hydrogen remaining, which would cause
a Gunn-Peterson effect at low redshifts. The observed lack of a
Gunn-Peterson effect by $z\approx 6$ is therefore a strong constraint
on preheating, or may imply the need for much higher escape fractions
at high redshifts.

An important result from this work is that no model consistent with
the $z\lsim 4$ temperature data produces sufficient entropy to explain
the high observed entropies in cluster cores. Furthermore, much of the
entropy which is injected into the IGM is lost through cooling soon
after preheating occurs.

Filtering masses computed from the volume averaged thermal history of
the IGM can reach values comparable to the mass of halos hosting $L_*$
galaxies today if preheating is particularly energetic or early. In
such cases we have shown that far too few $L_*$ and fainter galaxies
are produced, allowing us to rule out these models. However, we find
other preheated models which produce a galaxy luminosity function in
excellent agreement with the data, at least for faint magnitudes, and
without the need for supernovae feedback at late times. This comes at
the expense of over predicting the abundance of bright galaxies
however. When density-dependent filtering masses are used we find
that, in the dense regions of the IGM where galaxies are most likely
to form, rapid cooling after preheating keeps the filtering mass low
until late times, resulting in much less suppression of galaxy
formation. Consequently, more energetic preheating is required to
achieve the same degree of suppression in the luminosity function
compared to models with a globally averaged filtering mass. As a
result, we find no preheating model consistent with the IGM
temperature data that is able to fully match the observed luminosity
function at the faint end.

An interesting conclusion is that we find no model which is able to
adequately fit the luminosity function and is also consistent with the
observed lack of a Gunn-Peterson effect at $z\lsim 6$ for an escape
fraction $f_{\rm esc}=10\%$. Too much neutral hydrogen remains after
preheating, resulting in a large optical depth. This occurs as
preheating strongly suppresses galaxy formation, reducing the number
of ionizing photons produced below the number needed to fully ionize
the Universe. However, this small neutral fraction does allow the
models to potentially explain the Gunn-Peterson effect seen in Sloan
Digital Sky Survey quasar spectra at $z\approx 6$. This problem may be
alleviated by adopting a higher escape fraction (e.g. 20\% for $z_{\rm
preheat}=9$ and $E_{\rm preheat}=0.1$keV) without significantly
altering the thermal evolution of the IGM or the $z=0$ galaxy
luminosity function. Alternatively, if quasars are much more abundant
at $z>6$ than assumed in our calculations (which use the fitting
function of \pcite{mhr99}, which in turn is derived from observations
of quasars at $z<4.5$) they may provide sufficient photoionizations at
high redshift to adequately reduce the neutral hydrogen fraction.

Finally, we examined the abundances of satellite galaxies in the Local
Group. While preheating is able to flatten the predicted luminosity
function for relatively bright satellites---bringing it into agreement
with the observational data---it steepens the luminosity function at
faint magnitudes and so is unable to explain the paucity of the
faintest satellites.

Preheating alone can produce a galaxy luminosity function almost as
flat as that observed without the need for feedback from supernovae as
is commonly assumed in galaxy formation models. However, preheating
acting alone is not able to fully match the observational
constraints. The problem here is that when the faint end of the
luminosity function is significantly suppressed too much gas remains
available for cooling at late times. This gas is then able to cool in
massive halos, producing an overabundance of bright galaxies
(\pcite{benson03b}---see also \pcite{kauff99,cole00,sp}). In
conclusion, an early epoch of preheating has important consequences
for galaxy formation at recent times, and may remove or reduce the
need for more traditional forms of feedback in CDM models.

\section*{Acknowledgments}

AJB acknowledges the hospitality of the University of California at
Santa Cruz where part of this work was completed, and the Institute
for Computational Cosmology at the University of Durham whose
computing resources were used in parts of this work. We would like to
thank Carlton Baugh, Shaun Cole, Carlos Frenk and Cedric Lacey for
making available the {\sc galform} code for this work, and the
referee, Joop Schaye, for several valuable suggestions which improved
this paper. Support for this work was provided by NASA through grant
NAG5-11513 and by NSF grant AST-0205738 (PM).

\end{document}